\documentclass[12pt]{article}
\usepackage{cite}
\usepackage{amsmath,amssymb,amsfonts}
\usepackage{algorithmic}
\usepackage{graphicx}
\usepackage{textcomp}
\usepackage{xcolor}
\begin{document}

\title{Interferometric measurement of Van Hove singularities in strained graphene}

\author{Samad Roshan Entezar \\
University of Tabriz, Tabriz, Iran \\
s-roshan@tabrizu.ac.ir}

\maketitle

\begin{abstract}
This study presents a method based on the total internal reflection
and the phase-shifting interferometry for measuring the Van Hove
singularities in strained graphene. A linearly polarized light
passes through some quarter- and half-wave plates, a
hemi-cylindrical prism, and a Mach-Zehnder interferometer. The Van
Hove singularities manifest themselves as some sharp dips or peaks
in the spectrum of the final phase difference of the two
interference signals. The numerical analysis demonstrates that the
number of Van Hove singularities is independent of the modulus of
the applied stress, but their position shifts as the strength of the
tension increases. Besides, the number and location of singularities
strongly depend on the stress direction relative to the zigzag axis
in the graphene lattice. Moreover, we show that the location of
singularities is independent of the tension direction relative to
the tangential component of the electric field of the incident
radiation.
\\
keywords: Graphene, strain, phase-shifting interferometry, total
internal reflection.
\end{abstract}

\section{Introduction}

Graphene as a single sheet of carbon atoms in a hexagonal lattice
has shown abundant new physics and potential applications in
carbon-based nanoelectronic devices because of its exceptional
electrical and optical properties \cite{Novoselov2004Report,
Zhang2005Experimental, Geim2007The, Cresti2008Charge,
Castro2009The}. It has the highest known electrical and thermal
conductivity, as well as the highest stiffness and strength
\cite{Balandin2008Superior}. The ability to modify the electronic
structure of graphene is needed to make a graphene-based device or
circuit. Possible tools for achieving this goal include patterning,
electric field effects, and chemical doping
\cite{Novoselov2004Electric, Yan2007Intrinsic, Biel2009Anomalous}.
Recently, it has even been proposed that strain can be used to
achieve various essential elements for all-graphene electronics
\cite{Pereira2009Strain}. The strain effect in graphene provides a
new way to manipulate electron transport without external fields.
Graphene is structurally more tolerant than silicon to external
effects such as strain. Graphene, as the strongest material ever
measured, can also support strain well beyond the linear regime,
without to be bent or wrinkled. Very high strains can be easily
exerted on graphene before the mechanical failure
\cite{Lee2008Measurement, Ni2008Uniaxial, Mohiuddin2009Uniaxial,
Kim2009Large, Bao2009Controlled}. The effect of strain on graphene
has been studied both experimentally and theoretically
\cite{Gui2008Band, Pereira2009Tightbinding, Farjam2009Comment,
Gui2009Reply, Guinea2009Energy}. The mechanical strain often has
unexpected effects on the electronic properties of carbon
nanomaterials \cite{Heyd1997Uniaxial, Yang1999Band,
Tombler2000Reversible, Yang2000Electronic, Yu2000Strength,
Minot2003Tuning, Sazonova2004A}. Graphene offers a new opportunity
to discover such interesting electromechanical properties in two
dimensions \cite{Novoselov2004Electric}. Mechanically strained
graphene may introduce a new environment to study the novel physical
phenomena \cite{Novoselov2005Two, Park2008Electron, Park2008New}.
The strain may be applied naturally or intentionally to graphene.

In many applications, graphene lays or grows on a substrate. Growth
of graphene on substrates like SiO2 or SiC with a different lattice
constant usually introduces strain due to surface corrugations or
lattice mismatch \cite{Ferralis2008Evidence, Teague2009Evidence},
which can be detected by Raman spectroscopy
\cite{Borysiuk2009Transmission}. Even in the absence of a mismatch,
the strain still occurs along the edges and exhibits some
interesting quantum properties \cite{Jun2008Density,
Huang2009Quantum}. Bending the substrates may induce a uniaxial
strain on it \cite{Ni2008Uniaxial, Mohiuddin2009Uniaxial,
Kim2009Large, Huang2009Phonon}. Using tension to control the
physical properties of graphene has attracted an ever-increasing
interest \cite{Vozmediano2010Gauge, Novoselov2012A,
Bissett2014Strain, Roldan2015Strain, Barraza2015Discrete,
Galiotis2015Graphene, Jiang2015A, Amorim2016Novel, Deng2016Wrinkled,
Meunier2016Physical}.

From a technological point of view, understanding how strain affects
graphene's electronic and optical properties is of paramount
importance. Since the electronic states near the Fermi level are
directly related to optical and transport properties of graphene,
most researchers have focused on the electronic structure of
graphene near the Dirac points \cite{Partoens2006From,
Yamamoto2006Edge, Hsu2007Selection, Ezawa2007Metallic,
Heiskanen2008Electronic, Yin2010Coherent, Hancock2010Generalized,
Zarenia2011Energy}. Moreover, the density state may exhibit one or
more Van Hove singularities. The wavevectors at which Van Hove
singularities occur are often referred to as critical points of the
Brillouin zone \cite{Hove1953The}. In bulk solids, the presence of
Van Hove singularities in the density of states will result in
strong optical absorption peaks as well as Raman signals. However,
in graphene systems, few empirical studies show the existence of Van
Hove singularities \cite{Hao2009Optical, Yang2009Excitonic}. The
existence of Van Hove singularities in graphene structures may be
influenced by strain \cite{Pereira2010Optical}. Motivated by this,
we want to investigate Van Hove singularities in strained graphene.
In this paper, we present a method for measuring the frequency of
Van Hove singularities, based on Fresnel equations
\cite{Born1999Principles} and the phase-shifting interferometry
\cite{Malacara2007Optical}. A linearly polarized light beam passes
through a phase-sensitive internal reflection device, consisting of
a hemi-cylindrical prism, two half-wave plates, and two quarter-wave
with proper azimuth angles. The output beam enters a Mach-Zehnder
interferometer with two acousto-optic modulators separately placed
in two arms. Eventually, the two linearly orthogonally polarizations
of each output beam of the interferometer interfere with each other
as they pass through two analyzers, respectively. The ultimate phase
difference between the two interference signals is associated with
the azimuth angle of the fast axis of the half-wave plates, yielding
the sensitivity-tunable functionality.

\subsection{Theoretical calculation and model}

\begin{figure}[b]
\centering
\includegraphics[width=.5\textwidth]{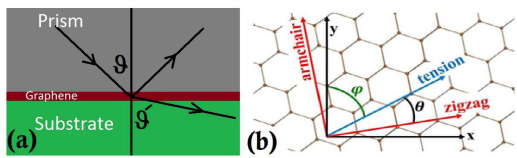}
\caption{a) The geometry of the problem with monolayer graphene
deposited on a substrate. Light is incident from a high-index prism
at the incidence angle $\vartheta$. b) The graphene honeycomb
lattice with its zigzag and armchair directions, and the tension
direction. Here, $\theta$ and $\varphi$ denote the angles which the
applied strain makes with the zigzag path and the tangential
electric field, respectively.}\label{fig1}
\end{figure}
Let us consider a single flat sheet of mono-layer graphene deposited
on a substrate of refractive index $n_s$, which is capped by a
high-index prism of refractive index $n_p$. A laser beam is incident
upon the base of the prism at the angle of incidence $\vartheta$
(see Fig. \ref{fig1}(a)). The reflection and transmission
coefficients of the structure are determined by satisfying the
continuity of the tangential component of the electric fields and
the discontinuity of the tangential component of the magnetic fields
caused by the induced surface current in graphene as
\begin{eqnarray}
  r_s &=& \frac{n_p\cos \vartheta -n_s \cos
\acute{\vartheta}-\eta_0\sigma}{n_p\cos \vartheta +n_s \cos
\acute{\vartheta}+\eta_0\sigma},\nonumber \\
  r_p &=& \frac{n_s\cos \vartheta -n_p
\cos \acute{\vartheta}+\eta_0\sigma \cos \vartheta \cos
\acute{\vartheta}}{n_s\cos \vartheta +n_p \cos
\acute{\vartheta}+\eta_0\sigma \cos \vartheta \cos
\acute{\vartheta}},
\end{eqnarray}
\begin{eqnarray}
  t_s &=& frac{2n_p\cos \vartheta}{n_p\cos \vartheta +n_s \cos
\acute{\vartheta}+\eta_0\sigma}, \nonumber \\
  r_p &=& \frac{2n_s\cos
\vartheta}{n_s\cos \vartheta +n_p \cos
\acute{\vartheta}+\eta_0\sigma \cos \vartheta \cos
\acute{\vartheta}}.
\end{eqnarray}
\begin{figure}[t]
\centering
\includegraphics[width=.4\textwidth]{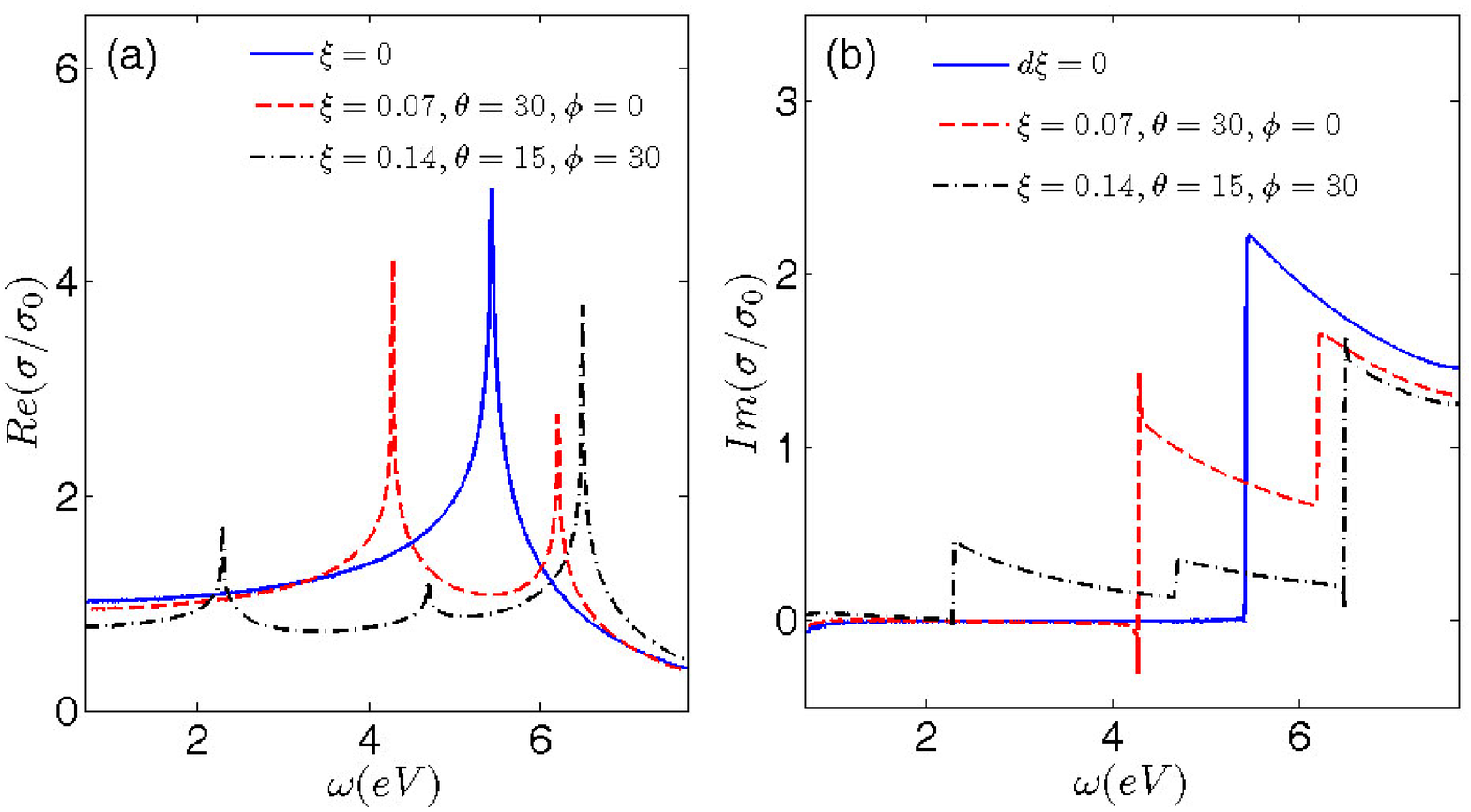}
\caption{a) The real and b) the imaginary parts of the normalized
optical conductivity of the strained graphene versus photon energy
$\omega$.}\label{fig2}
\end{figure}
Here, $\eta_0$ is the intrinsic impedance of free space, and
$\sigma$ is the optical conductivity of graphene monolayer. As one
knows, the optical conductivity of the graphene is the sum of the
intraband conductivity (due to the intraband electron-photon
scattering processes) and the interband conductivity (which
originates from the direct interband electron transitions)
\cite{Falkovsky2007Optical}. The optical conductivity of the
graphene has been derived using perturbation theory in references
\cite{Brun2013Electronic, Pedersen2003Analytic}. It was shown that
the applied tension deforms the graphene lattice and distorts the
reciprocal lattice \cite{Pellegrino2010Strain}. Since the conduction
and valence energy bands of the graphene vary due to the lattice
deformation, the induced tension changes the optical conductivity of
graphene \cite{Pellegrino2010Strain}. We use the procedure explained
in \cite{Falkovsky2007Optical, Brun2013Electronic,
Pedersen2003Analytic, Pellegrino2010Strain} to obtain $\sigma$ under
the applied tension without giving its explicit form here. Figure
\ref{fig2} shows a) the real and b) the imaginary parts of $\sigma$
(in the unite of $\sigma_0=\frac{e^2}{4\hbar}$) versus the photon
energy ($\omega$) for applied tension with different modules at
different directions. Here, we assume that the direction of the
applied tension makes angles $\theta$ and $\varphi$ with the zigzag
path in the graphene honeycomb lattice and the tangential electric
field of the incident beam respectively, and $\zeta$ indicates the
strain modulus (see Fig. \ref{fig1}(b)). We see some maxima at the
real part of $\sigma$ which, are attributed to the Van Hove
singularities at the saddle-points of the electronic band structure
of the graphene in the UV band \cite{Pereira2010Optical,
Pellegrino2010Strain, Pedersen2011Tight}. In unstrained graphene,
there are three equivalent saddle-points in the electronic band
structure, which show themselves as a single peak in the optical
conductivity (see dotted lines in Fig. \ref{fig2}). In the strained
graphene, at most two of the saddle-points may be equivalent. As a
result, we see two or three peaks in $\sigma$. The number of peaks
and their position strongly depend on the modulus and direction of
the applied tension.

\begin{figure}[t]
\centering
\includegraphics[width=.6\textwidth]{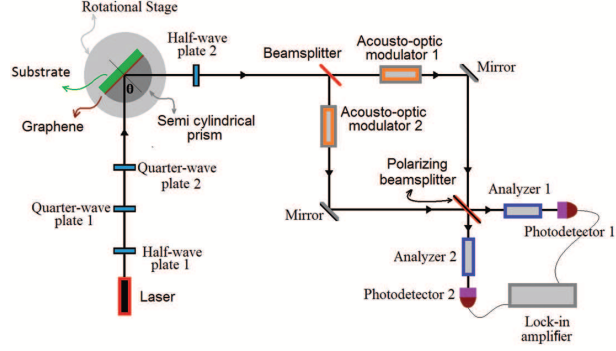}
\caption{Schematic diagram for measuring the phase difference
$\psi$.}\label{fig3}
\end{figure}
Here, we would like to present a simple procedure to obtain the
effect of applied tension on the position of the Van Hove
singularities. The schematic experimental setup for this measurement
is shown in Fig. \ref{fig3}. For convenience, we consider the
direction of the propagated light beam as the $z$-axis and set the
$x$-axis perpendicular to the plane of the paper. A linear
$s$-polarized laser light after passing through the half-wave plate
$1$ (with the fast axis at $\Delta/2$ with respect to the x-axis),
the quarter-wave plate $1$ (with the slow axes at $45^\circ$ with
respect to the x-axis) and the quarter-wave plate $2$ (with its slow
axes along the x-axis) is incident vertically on a hemi-cylindrical
prism with the refractive index of $n_p$. The light ray is refracted
into the prism, and it propagates at the incidence angle $\vartheta$
toward the boundary surface between the prism and the graphene
monolayer. For $\vartheta$ larger than the critical angle
$\sin^{-1}\frac{n_s}{n_p}$, the light undergoes total internal
reflection at the boundary. Then the reflected light beam travels
through the half-wave plate $2$, which its fast axes makes angle
$\alpha$ with the x-axis. The Jones vector of the light after the
half-wave plate $2$ can be written as:
\begin{eqnarray}
  E_t &=& \left(%
\begin{array}{c}
 A_se^{i\phi_s} \\
 A_pe^{i\phi_p} \\
\end{array}%
\right) \nonumber \\
   &=& \frac{t \acute{t}}{\sqrt{2}}\left(%
\begin{array}{c}
 r_s\cos 2\alpha e^{-i\Delta}+r_p \sin 2\alpha e^{-i\Delta} \\
 r_s\sin 2\alpha e^{-i\Delta}-r_p \cos 2\alpha e^{-i\Delta} \\
\end{array}%
\right),
\end{eqnarray}
where
\begin{equation}
t=\frac{2}{n_p+1},~\acute{t}=\frac{2n_p}{n_p+1},
\end{equation}
Here, $t$, $\acute{t}$ are the transmission coefficients at the
air-prism and the prism-air interface, respectively. A $50/50$ beam
splitter of a Mach-Zehnder interferometer splits the reflected light
beam into reflected and transmitted beams after passing through the
half-wave plate $2$. Those beams pass through the electro-optical
modulators $1$ and $2$ to shift their frequencies to $\omega_1$ and
$\omega_2$, respectively. Then, a polarizing beam splitter
superimposes the beams. The output beams after passing through the
analyzers $1$ and $2$ (which their transmission axis makes angle
$\beta$ with the x-axis) are detected using the photo-detectors $1$
and $2$ in both output ports. The intensities measured by the
photo-detectors $1$ and $2$ are
\begin{eqnarray}
  I_1&=&\frac{1}{2}(A_s^2 \cos^2 \beta + A_p^2 \sin^2 \beta \\
   &+& \sin 2\beta
A_s A_p \cos[(\omega_2-\omega_1)t+(\phi_p-\phi_s)]),\nonumber
\end{eqnarray}
\begin{eqnarray}
  I_2&=&\frac{1}{2}(A_s^2 \cos^2 \beta + A_p^2 \sin^2 \beta \\
   &+& \sin 2\beta
A_s A_p \cos[(\omega_2-\omega_1)t-(\phi_p-\phi_s)]). \nonumber
\end{eqnarray}
The intensities $I_1$ and $I_2$, are sent to a lock-in amplifier to
obtain the phase difference $\psi=2(\phi_p-\phi_s)$.

\section{Results and numerical calculations}

\begin{figure}[t]
\centering
\includegraphics[width=.5\textwidth]{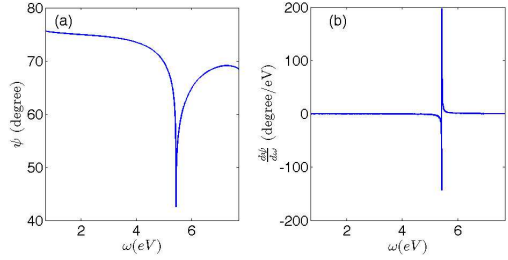}
\caption{a) Phase difference $\psi$ and b) the sensitivity
$\frac{d\psi}{d\omega}$ vs photon energy ($\omega$) in the absence
of the strain.}\label{fig4}
\end{figure}

In this section we describe the numerical analysis. Here, we assumed
that the incidence angle $\vartheta$ to be $\approx52.2^\circ$,
$n_p=1.7786$, $n_s=1.4$, $\alpha\cong25^\circ$,
$\Delta\approx25^\circ$ and $\beta\approx0^\circ$, respectively.
Fig. \ref{fig4} depicts the simulated results by plotting a) the
phase difference $\psi$ and b) the sensitivity
$\frac{d\psi}{d\omega}$ vs photon energy ($\omega$) in the absence
of the strain. As it is clear from Fig. \ref{fig4}(a), the phase
difference $\psi$ monotonically decreases to reach its minimum at a
specific frequency and then increases by increasing $\omega$. We
find that the phase difference changes its behavior at the frequency
in which the Van Hove singularity occurred (compare Figs.
\ref{fig4}(a) and \ref{fig2}(a) (the dotted-dashed line)). In this
method, the Van Hove singularity exhibit itself as a sharp peak or
deep valley in the phase difference curve. From Fig. \ref{fig4}(b),
we find that this method has an acceptable sensitivity
($\frac{d\psi}{d\omega}$) to obtain the frequencies at which the Van
Hove singularities occur.
\begin{figure}[h b]
\centering
\includegraphics[width=.3\textwidth]{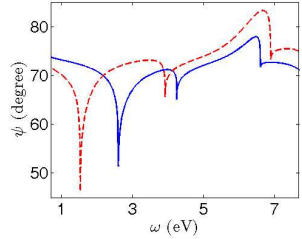}
\caption{The phase difference $\psi$ vs the photon energy ($\omega$)
for two different modului $\zeta=0.14$ (the solid line) and
$\zeta=0.21$ (the dashed line) of the applied strain with the fixed
direction at $\theta=20^\circ$, and $\varphi=0$.}\label{fig5}
\end{figure}
Now we consider the phase difference curve for the case of strained
graphene. We first study the case of graphene under strain in a
given direction with different moduli of applied strain. Figure
\ref{fig5} shows the phase difference $\psi$ vs the photon energy
($\omega$) for two different moduli $\zeta=0.14$ (the solid line)
and $\zeta=0.21$ (the dashed line) of the applied strain with the
fixed direction of the applied strain at $\theta=20^\circ$ and
$\varphi=0$. From the figure, we see that the number of sharp deeps
in the phase difference curve has increased to three due to the
stress, which indicates the existence of three Van Hove
singularities in the density of states of the strained graphene. It
is clear from the figure that the position of the singularities
changes slightly as the stress increases. The first two
singularities move to lower frequencies, and the third singularity
suffers from a blue-shift. We also note that the number of Van Hove
singularities is independent of the modulus of applied stress.

\begin{figure}[t]
\centering
\includegraphics[width=.5\textwidth]{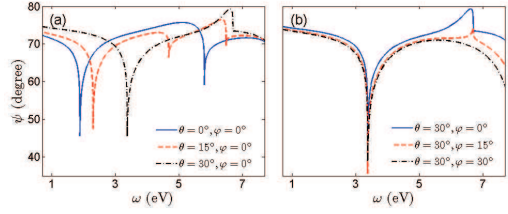}
\caption{The phase difference $\psi$ vs the photon energy ($\omega$)
for three different directions of the applied strain at a)
$\theta=0^\circ$, (the solid line), $\theta=15^\circ$, (the dashed
line), $\theta=30^\circ$ with $\varphi=0$ (the dotted-dash line) and
b) $\varphi=0^\circ$ (the solid line), $\varphi=15^\circ$,
$\varphi=30^\circ$ (the dashed line)$ with \theta=30^\circ$. Here,
the modulus of the applied strain ($\zeta$) is $0.14$.}\label{fig6}
\end{figure}
Finally, we consider the phase difference curve for the case of
strained graphene at different directions of the applied strain with
a given modulus. Figure \ref{fig6} depicts the phase difference
$\psi$ vs the photon energy ($\omega$) for three different
directions of the applied strain at a) $\theta=0^\circ$, (the solid
line), $\theta=15^\circ$, (the dashed line), $\theta=30^\circ$ (the
dotted-dash line) with $\varphi=0$ and b) $\varphi=0^\circ$ (the
solid line), $\varphi=15^\circ$, $\varphi=30^\circ$ (the dashed
line) with $\theta=30^\circ$. Here, we considered the modulus of the
applied strain ($\zeta$) to be $0.14$. As one knows, $\theta$ is the
angle between the direction of the applied tension and the zigzag
path in the graphene honeycomb lattice, and $\varphi$ indicates the
angle between the directions of the applied tension and the
tangential electric field of the incident beam. Figure \ref{fig6}
reveals that the number of the Van Hove singularities strongly
depends on both $\theta$ and $\varphi$, and it varies from one to
three. However, the frequency location of the Van Hove singularities
depends solely on $\theta$ and it is independent of $\varphi$.

\section{Conclusions}

 An experimental method for measuring the frequency of the Van Hove
singularities in the density of states of strained graphene is
proposed. The phase difference of p- and s-polarized reflected light
under the total internal reflection in a hemi-cylindrical prism,
whose base is contacted with the strained graphene, can be measured
accurately with the phase-shifting interferometry. The Van Hove
singularities present themselves as some sharp dips or peaks in the
phase difference spectrum. The validity of the method has been
examined using the special equations derived to estimate the phase
difference of the interferometric signals. We showed that the number
of Van Hove singularities strongly depends on the direction of the
applied strain. Not only the direction of the applied strain but
also its modulus affects the frequency locations of the Van Hove
singularities. We demonstrated the phase-shifting interferometry has
an acceptable sensitivity to obtain the frequency of the Van Hole
singularities.

\end{document}